\documentclass[5p,times,twocolumn]{elsarticle}
\usepackage{amsmath,amssymb}
\biboptions{comma,sort&compress}
\usepackage[utf8]{inputenc}
\usepackage{graphicx}
\usepackage{bm}
\usepackage[dvipsnames]{xcolor}
\usepackage{hyperref} 
\hypersetup{colorlinks=true, linkcolor=blue, anchorcolor=red, citecolor=blue, filecolor=red, urlcolor=red, pdfauthor=author}

\newcommand{\trento}{\texttt{T$_\mathrm{R}$ENTo}}

\newcommand{\SMASH}{\texttt{SMASH}}
\newcommand{\music}{\texttt{MUSIC}}
\newcommand{\is}{\texttt{iS3D}}
\newcommand{\sqrts}{\sqrt{s_\textrm{NN}}}
\newcommand{\Tsw}{T_{\text{sw}}}

\graphicspath{{fig/}}

\usepackage{cuted}   
\usepackage{placeins}
\usepackage{caption} 

\journal{Physics Letters B}

\begin{document}

\begin{frontmatter}

\title{Phenomenological constraints on QCD transport with quantified theory uncertainties}

\author{Sunil Jaiswal}
\ead{jaiswal.61@osu.edu}
\address{Department of Physics, The Ohio State University, Columbus, Ohio 43210, USA}

\date{\today}

\begin{abstract}
We present data-driven, state-of-the-art constraints on the temperature-dependent specific shear and bulk viscosities of the quark–gluon plasma from Pb--Pb collisions at $\sqrt{s_{\mathrm{NN}}}=2.76\,\mathrm{TeV}$. We perform global Bayesian calibration using the JETSCAPE multistage framework with two particlization ans\"atze, Grad 14-moment and first-order Chapman–Enskog, and quantify theoretical uncertainties via a centrality-dependent model discrepancy term. When theoretical uncertainties are neglected, the specific bulk viscosity and some model parameters inferred using the two ans\"atze exhibit clear tension. Once theoretical uncertainties are quantified, the Grad and Chapman–Enskog posteriors for all model parameters become almost statistically indistinguishable and yield reliable, uncertainty-aware constraints. Furthermore, the learned discrepancy identifies where each model falls short for specific observables and centrality classes, providing insight into model limitations.
\end{abstract}

\end{frontmatter}


\section{Introduction} \label{sec:intro}
High-energy nucleus–nucleus collisions at the Relativistic Heavy Ion Collider and the Large Hadron Collider create short-lived droplets of quark–gluon plasma (QGP), a deconfined state of quarks and gluons, enabling detailed studies of quantum chromodynamics (QCD) and QGP properties~\cite{Busza:2018rrf, Arslandok:2023utm, Achenbach:2023pba}. One of the primary goals of heavy-ion physics is to determine the equation of state and transport coefficients of the strongly coupled QGP. Although the equation of state of deconfined nuclear matter at vanishing baryon chemical potential is now well constrained by lattice QCD calculations~\cite{Borsanyi:2013bia, Bazavov:2014pvz}, determining transport coefficients from first principles remains challenging. Lattice extractions require reconstructing real-time spectral functions from Euclidean energy–momentum correlators, an ill-posed analytic continuation that introduces systematic and numerical uncertainties~\cite{Bazavov:2019lgz}. Moreover, in the experimentally accessible temperature range ${\sim\,}150{-}350$\,MeV, the strong coupling of the medium hinders perturbative approaches~\cite{Ghiglieri:2018dib, Shuryak:2019ydl}.

A data-driven approach aims to constrain the transport properties of QCD matter by comparing predictions from multistage heavy-ion collision models with collider measurements. Several Bayesian analyses have been performed, providing constraints on the transport coefficients~\cite{Novak:2013bqa, Sangaline:2015isa, Bernhard:2015hxa, Bernhard:2016tnd, Bernhard:2019bmu, Yang:2020oig, JETSCAPE:2020shq, JETSCAPE:2020mzn, Nijs:2020ors, Nijs:2020roc, Liyanage:2023nds, Heffernan:2023gye, Heffernan:2023utr, Gotz:2025wnv}. However, tensions have been observed between studies employing different models and/or experimental observables. The primary reason for such discrepancies is the lack of accounting for theory uncertainties arising from imperfections in the models within the multistage heavy-ion framework. Without accounting for these uncertainties, inference can force the approximate model to reproduce the data, effectively absorbing model inadequacy into the inferred parameters and yielding posteriors that need not reflect their physical values. Quantifying these uncertainties is therefore essential for reliable parameter estimation.

In this Letter, we address this issue using a recently developed Bayesian framework that explicitly quantifies the theory uncertainty associated with model inadequacy~\cite{Jaiswal:2025hyp}.\footnote{
    This source of uncertainty is often termed model discrepancy~\cite{Kennedy2002, Higdon2004, Arendt:etal2012:1, Brynjarsdottir_2014}.}
The framework statistically models this uncertainty at the level of final observables as additive corrections to multistage-model predictions, informed by qualitative knowledge of the models' varying reliability across centrality classes. We show that this treatment removes tensions between parameter posteriors obtained within the JETSCAPE multistage model using Grad 14-moment~\cite{Grad} or first-order Chapman–Enskog (CE)~\cite{chapman1990mathematical, ANDERSON1974466, Jaiswal:2014isa} viscous corrections at particlization. With the theory uncertainty quantified, the inference no longer forces the model to fit the data, yielding mutually consistent and reliable constraints on the QGP viscosities.

\section{Data and multistage heavy-ion model} \label{sec:model}
We consider a broad set of centrality-binned, $p_T$-integrated measurements from Pb--Pb collisions at $\sqrts{}=2.76\,\mathrm{TeV}$: (i) the number of charged hadrons per unit pseudorapidity $dN_{\text{ch}}/d\eta$~\cite{Aamodt:2010cz}; (ii) the transverse energy per unit pseudorapidity $dE_T/d\eta$~\cite{Adam:2016thv}; (iii) the number of identified charged hadrons per unit rapidity $dN_i/dy,\ i\in \{\pi, k, p\}$~\cite{Abelev:2013vea}; (iv) the mean transverse momenta of identified hadrons $\langle p_T \rangle_i,\ i\in \{\pi, k, p\}$~\cite{Abelev:2013vea}; (v) the two-particle cumulant flow coefficients $v_n\{2\}$ for $n=2,3,4$ \cite{ALICE:2011ab}; (vi) the fluctuation in the mean transverse momentum $\delta p_T / \langle p_T \rangle$ \cite{Abelev:2014ckr}. In total, 110 observations across these 12 observables constitute a diverse dataset measured with adequate statistical accuracy.

The multistage simulation model considered in this work comprises the following modules, each describing a distinct stage of the system's evolution~\cite{JETSCAPE:2020shq, JETSCAPE:2020mzn}:
\begin{enumerate}
    \item \trento{}: A phenomenological model that simulates the initial energy deposition following the collision of the nuclei~\cite{Moreland:2014oya, trento_code}.

    \item Free-streaming: A weakly coupled pre-equilibrium stage modeling the system's evolution during the first $\sim$1~fm/$c$~\cite{Liu:2015nwa, Broniowski:2008qk, fs_code}.

    \item Relativistic viscous hydrodynamics: Simulates the dissipative evolution of QCD matter using the \music{} code~\cite{Schenke:2010nt, Schenke:2010rr, Paquet:2015lta, 2000JCoPh.160..241K, hydro_code}. The specific shear and bulk viscosities are parametrized as~\cite{JETSCAPE:2020mzn}:
    \begin{align}
        \left(\frac{\eta}{s}\right)(T) =&\,(\eta/s)_\mathrm{kink} + \Theta(T - T_{\rm kink}) a_\mathrm{high} (T - T_{\rm kink}) 
        \nonumber \\ \label{etabys_param}
        &+ \Theta(T_{\rm kink} - T) a_\mathrm{low} (T - T_{\rm kink}),
        \\ \label{zetabys_param}
        \left(\frac{\zeta}{s}\right)(T) =&\, \frac{(\zeta/s )_{\max}\,\Lambda^2}{\Lambda^2 + \left( T-T_\zeta\right)^2},
    \end{align}
    where $\Lambda =\, w_{\zeta} \left[1 + \lambda_{\zeta}\, {\rm sign} \left(T{-}T_\zeta\right) \right]$.

    \item Particlization: This stage models the conversion of the QGP fluid into hadrons as the system cools below the critical temperature. Local hadron phase-space distributions are constructed from the ten components of the energy–momentum tensor evolved by viscous hydrodynamics. The lack of additional information on the microscopic dynamics underlying the out-of-equilibrium hydrodynamical evolution results in an irreducible ambiguity in the hadron phase-space distribution. We consider the two maximally different models of viscous corrections to the local equilibrium distribution function studied in Ref.~\cite{JETSCAPE:2020shq, JETSCAPE:2020mzn}: the Grad 14-moment approximation~\cite{Grad} and first-order Chapman–Enskog (CE) viscous corrections~\cite{chapman1990mathematical, ANDERSON1974466, Jaiswal:2014isa}. The \is{} sampler implements the particlization process~\cite{McNelis:2019auj, is3d_code}.

    \item Hadronic decays and rescatterings: These processes are simulated using Boltzmann kinetic transport with the \SMASH{} code~\cite{Weil:2016zrk, smash_code}.
\end{enumerate}
The simulation employs $17$ parameters, listed in Table~\ref{param_table}; see Ref.~\cite{JETSCAPE:2020mzn} for details.

\begin{table}[!tb]
    \begin{tabular}{p{110pt} p{33pt} p{73pt} }
        \hline
        \hline
        Parameter name & Symbol & [min., max.]\\
        \hline
        Normalization & $N$ & [10, 20]\\
        Generalized mean & $p$ & [$-0.7$, 0.7]\\
        Multiplicity fluctuation & $\sigma_k$ & [0.3, 2]\\
        Nucleon width & $w$ & [0.5, 1.5] fm\\
        Min. dist. btw. nucleons & $d_{\text{min}}^3$ & [0, 1.7$^3$] fm$^3$\\
        Free-streaming time scale & $\tau_R$ & [0.3, 2] fm/c\\
        Free-streaming energy dep. & $\alpha$ & [$-0.3$, 0.3]\\
        Temperature of $(\eta/s)$ kink & $T_{\rm kink}$ & [0.13, 0.3] GeV\\
        Low temp. slope of $(\eta/s)$ & $a_{\text{low}}$ & [$-2$, 1] GeV$^{-1}$\\
        High temp. slope of $(\eta/s)$ & $a_{\text{high}}$ & [$-1$, 2] GeV$^{-1}$\\
        $(\eta/s)$ at kink & $(\eta/s)_{\rm kink}$ & [0.01, 0.2]\\
        Maximum of $(\zeta/s)$ & $(\zeta/s)_{\text{max}}$ & [0.01, 0.2]\\
        Temperature of $(\zeta/s)$ peak & $T_{\zeta}$ & [0.12, 0.3] GeV\\
        Width of $(\zeta/s)$ peak & $w_{\zeta}$ & [0.025, 0.15] GeV\\
        Asymmetry of $(\zeta/s)$ peak & $\lambda_{\zeta}$ & [$-0.8$, 0.8]\\
        Shear relaxation time factor & $b_{\pi}$ & [2, 8] \\
        Particlization temperature & $\Tsw$ & [0.13, 0.165] GeV\\
        \hline
        \hline
    \end{tabular}
    \caption{A list of all 17 parameters in the multistage simulation model. All prior distributions are assumed to be uniform and nonzero within the range quoted and zero outside.}
    \label{param_table}
\end{table}

In order to perform Bayesian parameter inference for the multistage model using ALICE data, we require model predictions at many parameter values. Because these simulations are computationally intensive, we train fast model surrogates (Gaussian process emulators) using a new and improved method -- Automatic Kernel Selection Gaussian Process (AKSGP)~\cite{AKSGP}. The simulation data used to train the emulators are the same as in Ref.~\cite{JETSCAPE:2020mzn}.

\section{Model-data comparison} \label{sec:model_data_comp}
We begin by setting up the model discrepancy framework~\cite{Kennedy2002, Jaiswal:2025hyp}. The statistical model for the $i$th measurement of an observable $y_{\rm exp}$ is
\begin{equation}\label{eq:obs}
    y_{\rm exp}(x_i) = \xi(x_i) + \epsilon_i \,.
\end{equation}
Here $x_i$ denotes centrality in this work, with the index $i$ corresponding to a centrality bin. We assume independent errors $\epsilon_i$ for $y_{\rm exp}(x_i)$, with $\epsilon_i \sim \mathcal{N}(0,\sigma_i^2)$, where $\sigma_i$ is taken from the ALICE-reported statistical uncertainties.

We denote the predictions from the multistage JETSCAPE model for the observable $y_{\rm exp}(x)$ as $\eta_{\rm mod}(x, \bm{\theta})$, where $\bm{\theta}$ is the vector of true but unknown model parameters. We model the relationship between $\xi(x)$ and $\eta_{\rm mod}(x, \bm{\theta})$ as
\begin{equation}\label{eq:discrep}
    \xi(x) = \eta_{\rm mod}(x, \bm{\theta}) + \delta_{\rm MD}(x) \,,
\end{equation}
where $\delta_{\rm MD}(x)$ quantifies the discrepancy between the true value of the observable $\xi(x)$ and the JETSCAPE prediction $\eta_{\rm mod}(x,\bm{\theta})$ at centrality $x$. Since information about the ``true'' dynamics is available only through measured final-state observables, we define $\delta_{\rm MD}$ as an observable-level discrepancy of the full multistage model. Thus $\delta_{\rm MD}(x)$ quantifies the theoretical uncertainty arising from inadequacies of the multistage model in predicting the observables. Combining Eqs.~\eqref{eq:obs} and \eqref{eq:discrep}, the observation $y_{\rm exp}(x_i)$ can be expressed as
\begin{equation}\label{eq:obs_with_md}
    y_{\rm exp}(x_i) = \eta_{\rm mod}(x_i, \bm{\theta}) + \delta_{\rm MD}(x_i)  + \epsilon_i \,.
\end{equation}
If the true parameter values $\bm{\theta}^*$ were known \textit{a priori}, then the discrepancy between $y_{\rm exp}(x_i)$ and $\eta_{\rm mod}(x_i,\bm{\theta}^*)$ would quantify how well the multistage model describes the data (up to measurement noise) without compensating through parameter shifts. Conversely, if $\delta_{\rm MD}^*(x) \equiv \xi(x) - \eta_{\rm mod}(x,\bm{\theta}^*)$ were known exactly (e.g. from theoretical calculations), then using $\delta_{\rm MD}^*(x)$ in Eq.~\eqref{eq:obs_with_md} would, in principle, yield correct estimates of $\bm{\theta}$. In practice, neither $\bm{\theta}^*$ nor $\delta_{\rm MD}^*(x)$ is known, and both must be inferred from data simultaneously.

While $\delta_{\rm MD}(x)$ is usually not known precisely, qualitative information about this discrepancy is often available. The Bayesian model discrepancy framework presented in Ref.~\cite{Jaiswal:2025hyp} incorporates such qualitative knowledge into parameter inference by treating $\delta_{\rm MD}(x)$ statistically, modeling it with a Gaussian process (GP),
 \begin{equation}
    \delta_{\rm MD}(\cdot \mid\bm\phi) \sim {\rm GP} \left(\bm 0, K(\cdot,\cdot \mid\bm\phi) \right) \,,
 \end{equation}
where $K(\cdot,\cdot \mid\bm\phi)$ is the covariance kernel with hyperparameters $\bm\phi$. The GP represents a distribution over functions, with kernel $K(\cdot,\cdot \mid\bm\phi)$ governing their properties across centrality. The kernel choice encodes qualitative knowledge about the error of the multistage model and will be discussed shortly. We assign uniform (over a limited range, see Table~\ref{param_table}) prior distributions to the model parameters $\bm\theta$ and the GP hyperparameters $\bm\phi$. Bayesian inference based on Eq.~\eqref{eq:obs_with_md}, with $\delta_{\rm MD}(\cdot\mid\bm\phi)$, then updates these priors to posterior distributions conditioned on the observations, providing uncertainty-quantified estimates of $\bm \theta$ and the discrepancy $\delta_{\rm MD}(x)$.

\begin{figure}[t!]
    \centering
    \includegraphics[width=\linewidth]{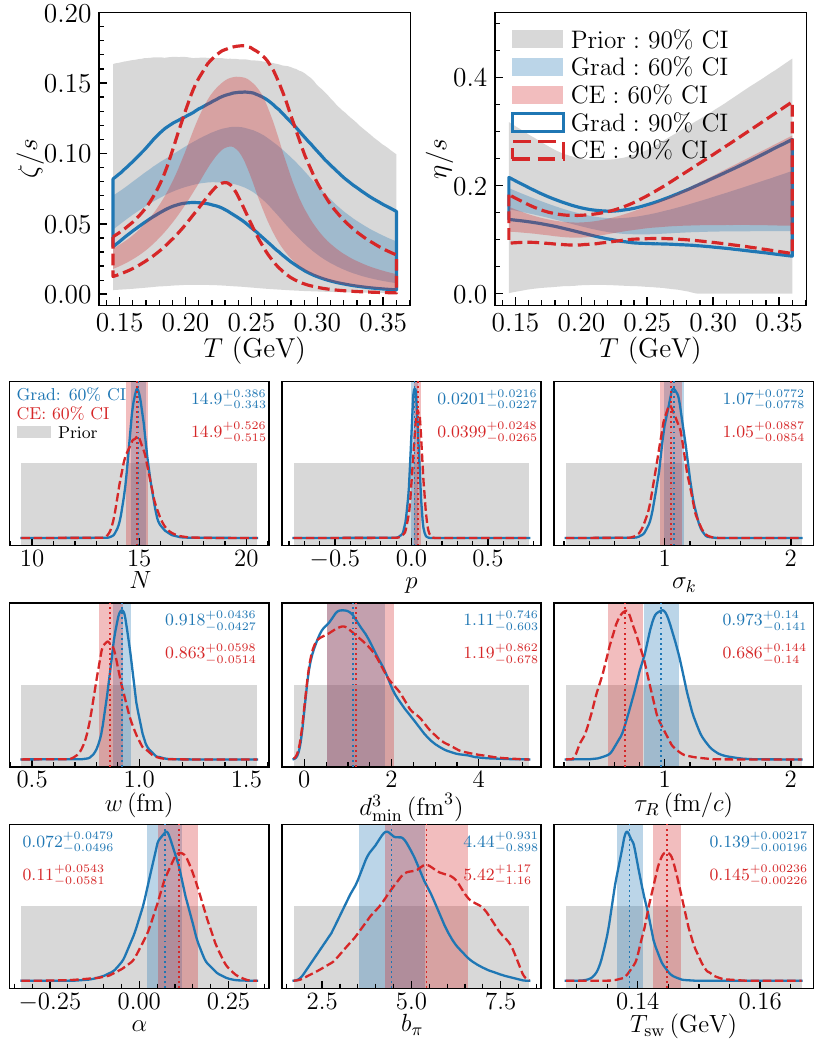}
    \vspace{-5mm}
    \caption{Parameter inference without accounting for model discrepancy (w/o MD). \textbf{Top row:} Posteriors for the specific bulk and shear viscosities, with central $90\%$ and $60\%$ credible interval (CI) bands. The shaded gray regions denote the $90\%$ prior intervals. \textbf{Bottom rows:} Solid and dashed curves show the posterior distributions for the remaining model parameters. Colored shaded regions show the central $60\%$ CI, and shaded gray regions indicate the uniform prior range. In each subplot, a dotted vertical line marks the posterior median, and the printed value reports $\text{median} \pm 60\%\, \mathrm{CI}$.}
    \vspace{-2mm}
    \label{fig:param_woMD}
\end{figure}

\begin{figure}[t!]
    \centering
    \includegraphics[width=\linewidth]{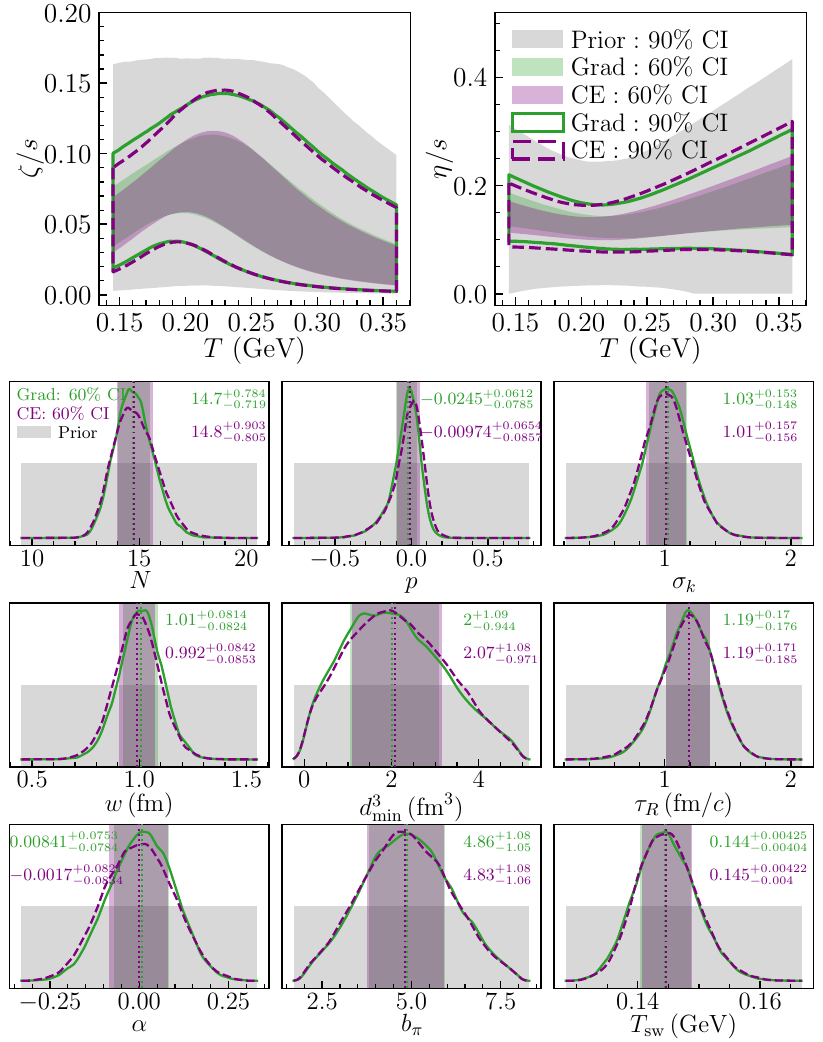}
    \vspace{-5mm}
    \caption{Parameter inference accounting for model discrepancy (w/ MD). The plot layouts and legends are identical to those in Fig.~\ref{fig:param_woMD}.}
    \vspace{-2mm}
    \label{fig:param_wMD}
\end{figure}

Careful studies have established that the hydrodynamic description at the core of the JETSCAPE multistage model improves with increasing opacity of the medium~\cite{Kurkela:2019kip, Ambrus:2024hks, Ambrus:2024eqa, Peng:2025gbj}. For a given collision system, this implies that the model's reliability degrades as one moves from the most central to more peripheral collisions. Also, since the Grad and Chapman–Enskog particlization schemes provide near-equilibrium corrections, they perform better in more central collisions. We therefore consider observable predictions from the JETSCAPE model to be more reliable at small centralities than at large ones, and encode this information by assigning each observable's $\delta_{\rm MD}$ a nonstationary GP kernel whose marginal variance is non-decreasing with centrality:
\begin{equation}\label{kernel1}
    K(x_i, x_j \mid\bm\phi) \equiv 
    s^2 + \bar{c}^2 (x_i x_j)^r \exp\left( -\frac{\lVert x_i-x_j\rVert^2}{2\ell^2} \right) ,\quad  r\geq 0,
\end{equation}
with $\boldsymbol{\phi}=\{s,\bar{c},r,\ell\}$. Here, $s$ sets a centrality-independent baseline, $\bar{c}$ the overall scale, $r$ controls how the variance grows with centrality ($r{=}0$ recovers a stationary kernel), and $\ell$ is the correlation length. With $s{=}\bar{c}{=}0$, the discrepancy $\delta_{\rm MD}$ vanishes. This construction allows the multistage model discrepancy to remain constant or increase with centrality (but not decrease), allowing for the expected degradation of model reliability toward peripheral collisions.


\begin{figure*}[t!]
    \centering
    \includegraphics[width=.938\linewidth]{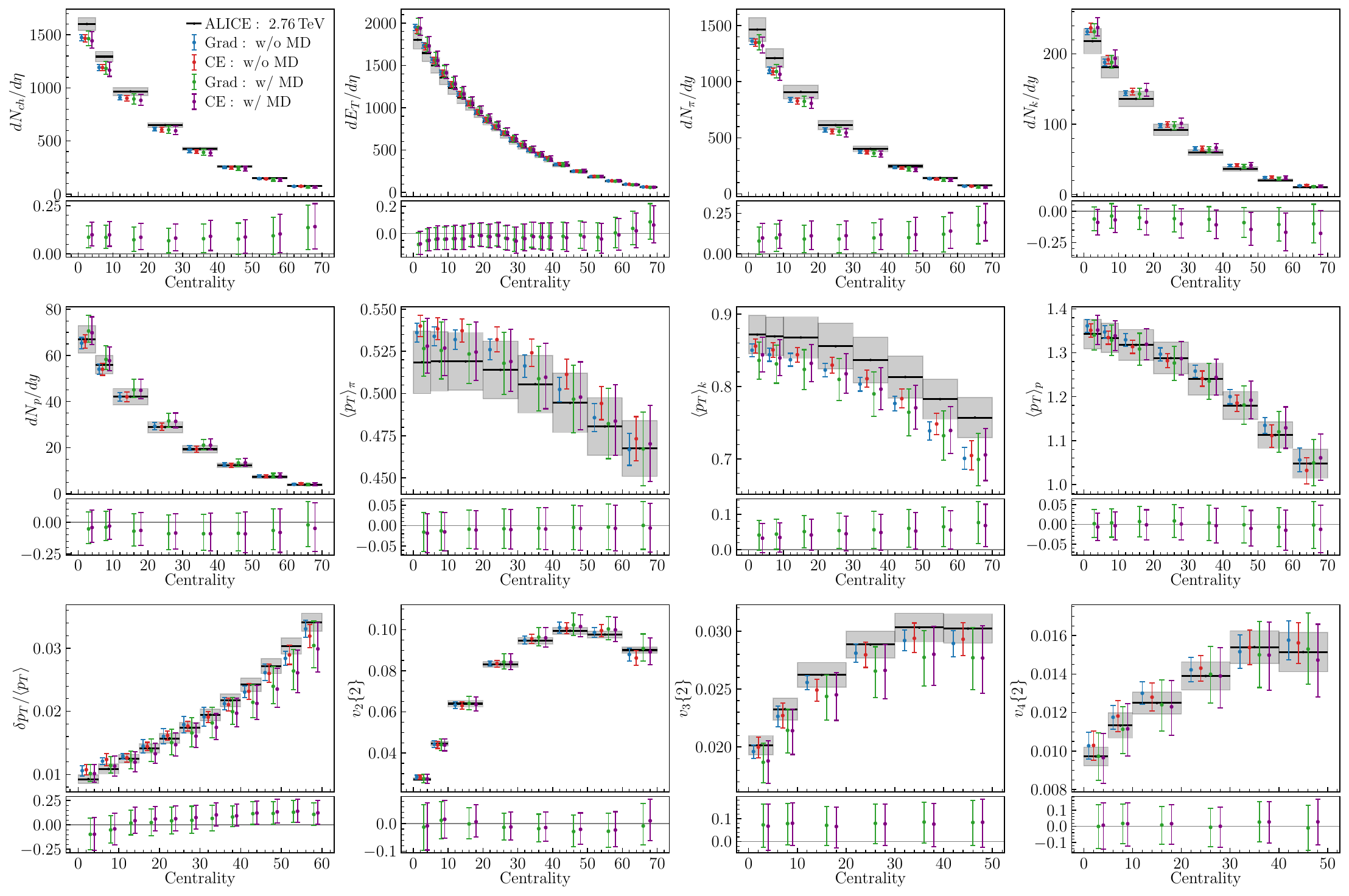}
    \vspace{-3mm}
    \caption{Model predictions, obtained using the parameter posteriors shown in Figs.~\ref{fig:param_woMD} and \ref{fig:param_wMD}, are compared with ALICE data (black). Panels below each observable show the normalized discrepancies $(y_{\rm exp}-\eta_{\rm mod})/|\langle y_{\rm exp}\rangle|$ from the w/ MD predictions.}
    \vspace{-3mm}
    \label{fig:obs_model}
\end{figure*}

We perform independent Bayesian inference for four cases: (i) using Grad particlization without model discrepancy (Grad: w/o MD); (ii) using Chapman–Enskog particlization without model discrepancy (CE: w/o MD); (iii) using Grad particlization with model discrepancy (Grad: w/ MD); (iv) using Chapman–Enskog particlization with model discrepancy (CE: w/ MD). Bayesian inference without MD involves 17 model parameters, whereas those with MD involve 65 parameters in total: 17 model parameters plus 4 GP hyperparameters per observable (12 observables $\Rightarrow$ 48 GP hyperparameters).

The parameter posteriors for both particlization methods without MD are shown in Fig.~\ref{fig:param_woMD}~%
    \footnote{These posteriors are tighter than those obtained in Refs.~\cite{JETSCAPE:2020shq, JETSCAPE:2020mzn} (which used the same ALICE data and JETSCAPE model) owing to the improved AKSGP model emulator employed here.}.
Significant tension can be observed between the specific bulk viscosity posteriors from Grad (blue) and CE (red) for all temperatures. The specific shear viscosity posteriors agree at high temperatures, but Grad predicts slightly larger values than CE below $\approx 200\,\mathrm{MeV}$. In the bottom rows of Fig.~\ref{fig:param_woMD}, several model parameters are seen to differ between the two calibrations. The particlization temperature is lower for Grad, $T_{\rm sw} = 138.65^{+2.17}_{-1.96}\,$MeV, than for CE, $144.79^{+2.36}_{-2.26}\,$MeV. Differences are also visible for the nucleon width $w$ in the \trento{} initial-condition model~\cite{Giacalone:2022hnz}, the free-streaming timescale $(\tau_R,\alpha)$, and the shear–relaxation-time factor $b_\pi$ in the hydrodynamic stage. This is an issue because, although only the particlization ansatz at freeze-out changes between the multistage models, the inferred parameters of the initial, pre-equilibrium, and hydrodynamic stages shift, indicating that model imperfections propagate across stages. These shifts indicate that model inadequacy is being absorbed by parameter shifts, compromising the parameters' physical interpretability.

\begin{figure*}[t!]
    \centering
    \includegraphics[width=\linewidth]{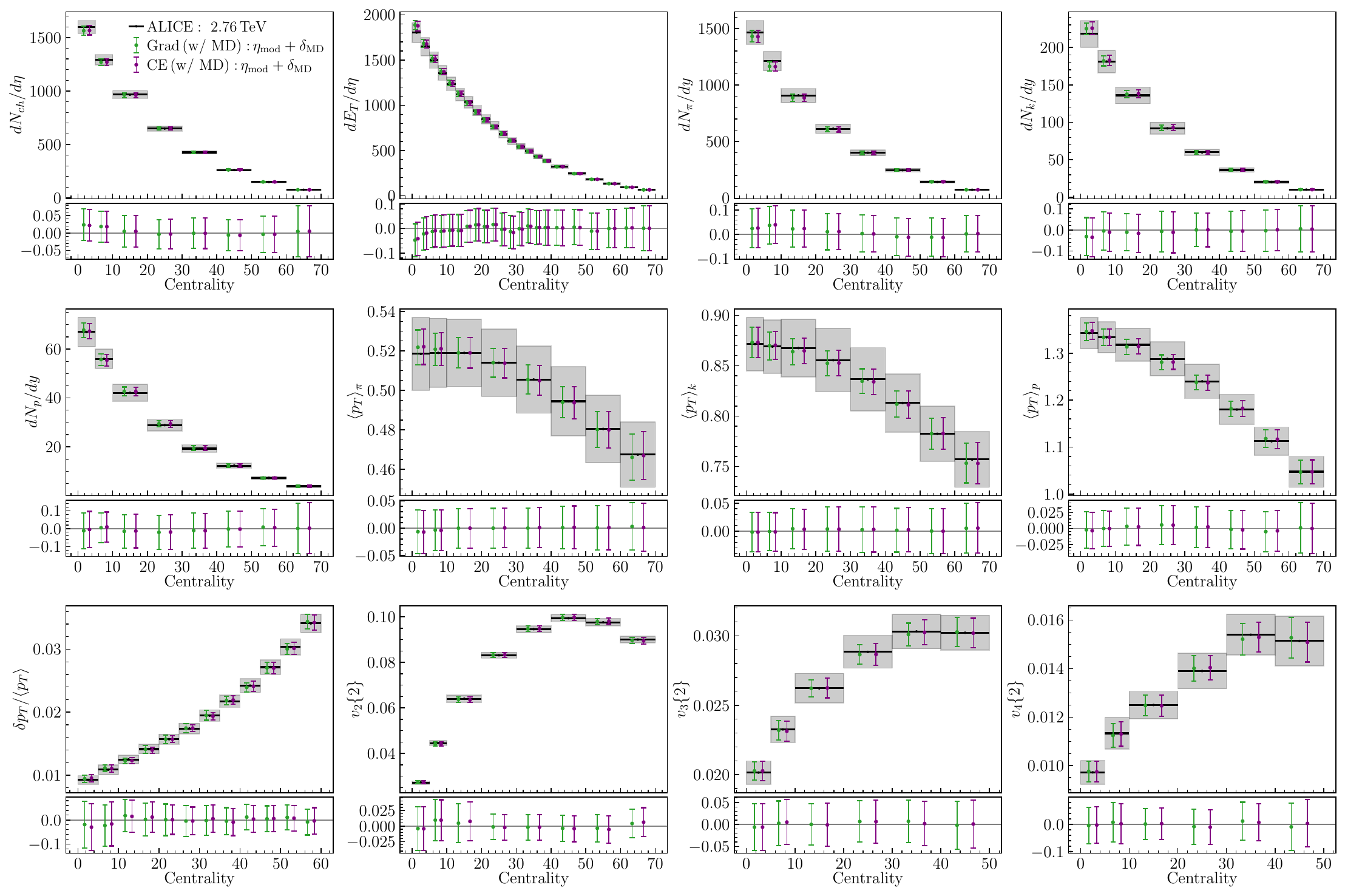}
    \vspace{-6mm}
    \caption{Observable predictions of $\eta_{\rm mod}+\delta_{\rm MD}$ obtained using the full parameter posteriors shown in Figs.~\ref{fig:param_woMD} and \ref{fig:param_wMD} together with the inferred hyperparameters of $\delta_{\rm MD}$ (not shown), compared with ALICE data. Panels below each observable show the normalized error $(y_{\rm exp}-\eta_{\rm mod}-\delta_{\rm MD})/|\langle y_{\rm exp}\rangle|$, with the horizontal line at $0$ indicating exact agreement with the mean experimental value.}
    \vspace{-2mm}
    \label{fig:obs_modelPlusDelta}
\end{figure*}

Figure~\ref{fig:param_wMD} shows the parameter posteriors for the uncertainty-quantified (w/ MD) cases. With quantified theory uncertainties, the Grad (green) and CE (purple) posteriors are statistically almost indistinguishable for $(\eta/s)(T)$, $(\zeta/s)(T)$, and all other model parameters; the tensions observed in w/o MD case disappear. Because the two calibrations use the same multistage model and differ only in the particlization ansatz, once the overall theory uncertainty of the multistage model is quantified, the inadequacy of each particlization scheme relative to the (unknown) true particlization is quantified as well, leading to convergence of the posteriors for all shared model parameters. As a result, shortcomings in the theoretical modeling of one stage do not propagate to other stages. This collapse of model dependence is the central result of this work, yielding the most reliable constraints on QGP viscosities to date.

In the top row of Fig.~\ref{fig:param_wMD} we observe an enhancement of the specific bulk viscosity around $T{\,\approx\,}180{-}220$\,MeV; small values are disfavored there. For the specific shear viscosity, the posterior favors values around $\approx 0.1{-}0.2$, with a modest increase towards both lower and higher temperatures away from $T\approx 220\,$MeV. Compared with those obtained in Ref.~\cite{JETSCAPE:2020shq} using Bayesian Model Averaging (BMA), the specific viscosity posteriors for $(\zeta/s)(T)$ and $(\eta/s)(T)$ in Fig.~\ref{fig:param_wMD} are statistically consistent, but more constrained.%
    \footnote{It should be noted that the specific-viscosity posteriors obtained using BMA in Ref.~\cite{JETSCAPE:2020shq} were calibrated using both Pb--Pb $2.76\,\mathrm{TeV}$ LHC data and Au--Au $200\,\mathrm{GeV}$ RHIC data, whereas the present work uses only Pb--Pb $2.76\,\mathrm{TeV}$ LHC data for calibration.}
We also see in the bottom row of Fig.~\ref{fig:param_wMD} that both Grad and CE now estimate consistent particlization temperatures $T_{\rm sw}$: $144.46^{+4.25}_{-4.04}\,\mathrm{MeV}$ (Grad) and $144.66^{+4.22}_{-4.00}\,\mathrm{MeV}$ (CE).

In Fig.~\ref{fig:obs_model} we compare the ALICE data used for calibration with posterior-predictive distributions for all four calibrations, constructed from the full joint posterior. Markers denote posterior medians and error bars show central $68\%$ credible intervals. The posterior-predictive distributions are obtained by drawing samples from the MCMC chain and, for each draw, sampling from the emulator predictive distribution to propagate emulator uncertainty (the emulator training accounts for simulation variance). Thus, the posterior-predictive construction propagates the relevant sources of uncertainty into the predictions, including emulator uncertainty, parameter uncertainty from calibration, and, in the w/MD analyses, uncertainty associated with the inferred model-discrepancy term. (For completeness, Fig.~\ref{fig:obs_modelPlusDelta} shows $(\eta_{\rm mod}+\delta_{\rm MD})$ predictions, which, as expected from Eq.~\eqref{eq:obs_with_md}, are in excellent agreement with the data.) In Fig.~\ref{fig:obs_model}, accounting for theory uncertainty enlarges the predictive intervals: the Grad (w/ MD) and CE (w/ MD) results have larger error bars than their w/o MD counterparts. This is because the parameter posteriors are wider (Fig.~\ref{fig:param_wMD} vs.~Fig.~\ref{fig:param_woMD}). Although the w/o MD predictions have smaller error bars, they are not reliable as the underlying parameter posteriors are not robust across particlization schemes (discussed above). Moreover, calibrating on a subset of ALICE data (Figs.~\ref{fig:param_woMD_subset} and \ref{fig:param_wMD_subset} in \ref{app:subset}) shifts several parameter posteriors significantly, indicating that the model is being forced to fit the calibration set by compensating through parameter shifts. This, in turn, leads to incorrect predictions for many observables not used in the calibration (Fig.~\ref{fig:obs_model_subset}).

The narrow panels below each observable in Fig.~\ref{fig:obs_model} show the normalized discrepancy $(y_{\rm exp}-\eta_{\rm mod})/|\langle y_{\rm exp}\rangle|$ for w/ MD predictions.%
    \footnote{For a sufficiently flexible discrepancy GP this coincides with the normalized learned discrepancy $\delta_{\rm MD}/|\langle y_{\rm exp}\rangle|$. In this study the two are nearly identical, since the discrepancy-corrected predictions $\eta_{\rm mod}+\delta_{\rm MD}$ closely follow the data (see Fig.~\ref{fig:obs_modelPlusDelta}).}
Values near zero indicate a good description, while deviations imply deficiencies in the model's description of the data. The Grad and CE residuals are similar, with their predictions agreeing with the data within uncertainties. No clear preference for one scheme over the other can be observed. For the w/ MD cases, we observe that $\langle p_T\rangle_\pi$, $\langle p_T\rangle_p$, $v_2\{2\}$, and $v_4\{2\}$ are well described. In contrast, systematic deviations from the data are observed for $dN_{\text{ch}}/d\eta$, $dN_\pi/dy$, $dN_k/dy$, $\langle p_T\rangle_k$, $\delta p_T/\langle p_T \rangle$, and $v_3\{2\}$, indicating sizable errors in the model predictions, $\delta_{\rm MD}$. The particularly large deviations in $v_3\{2\}$ and $\delta p_T/\langle p_T \rangle$ primarily suggest deficiencies in the initial-condition model. A detailed analysis of model performance and the sensitivity of model predictions to parameters will be reported separately. 

\section{Summary} \label{sec:summary}
\vspace{-.1cm}
Many problems in modern science rely on multistage or multi-model frameworks with parameters describing key properties of the system that must be inferred from data. Model–data comparison therefore requires careful quantification of both experimental and theoretical uncertainties. When models are used outside their domains of validity without accounting for theory uncertainties, parameter estimates can be biased, reducing model parameters to mere fitting variables~\cite{Kennedy2002, Higdon2004, Arendt:etal2012:1, Brynjarsdottir_2014}. Despite its importance, quantifying theoretical uncertainties remains challenging across disciplines -- e.g., cosmology~\cite{Mandelbaum:2017jpr}, astrophysics~\cite{Conroy:2013if}, systems biology~\cite{Gutenkunst:2007}, computational chemistry~\cite{Simm_2017, Medford_2014}, and climate/environmental modeling~\cite{Murphy2004}.

In this work, we employ a Bayesian framework that quantifies theoretical uncertainties (model discrepancy)~\cite{Jaiswal:2025hyp} to constrain the temperature-dependent shear and bulk viscosities of the quark-gluon plasma from Pb--Pb collision data at $\sqrt{s_{\mathrm{NN}}}=2.76\,\mathrm{TeV}$. We show that, without accounting for theory uncertainties, the multistage JETSCAPE framework with Grad and Chapman–Enskog particlization leads to tensions in the inferred transport coefficients and model parameters. Once the theoretical uncertainties are quantified, these tensions disappear and the Grad and CE posteriors become statistically indistinguishable across all shared parameters. This collapse of the model dependence yields reliable constraints on QGP viscosities in the temperature range $T{\sim\,}150{-}350\,$MeV, where first-principles calculations remain limited. The learned discrepancy further provides insight into the limitations of the theoretical models.

This work advances efforts to resolve long-standing model dependence in heavy-ion parameter extraction. The approach enables rigorous, uncertainty quantified Bayesian model–data comparison across heavy-ion physics and other disciplines. Open-source code is provided to facilitate adoption~\cite{jaiswal_2025_17186734}.

\vspace{-.2cm}
\section*{Acknowledgements}
\vspace{-.2cm}
I am grateful to Richard Furnstahl and Ulrich Heinz for useful feedback and edits that improved the manuscript's clarity. I also thank Chun Shen for useful comments on the manuscript. This research was supported by the CSSI program Award OAC-2004601 (BAND collaboration \cite{BAND_Framework}). Bayesian inference was performed on resources provided by the Ohio Supercomputer Center \cite{OhioSupercomputerCenter1987} (Project No.~PAS0254).


\appendix

\vspace{-.3cm}
\section{Inference with subset of observables} \label{app:subset}
\vspace{-.2cm}

\begin{strip}
\centering
\begin{minipage}[t]{0.47\textwidth}
  \centering
  \includegraphics[width=\linewidth]{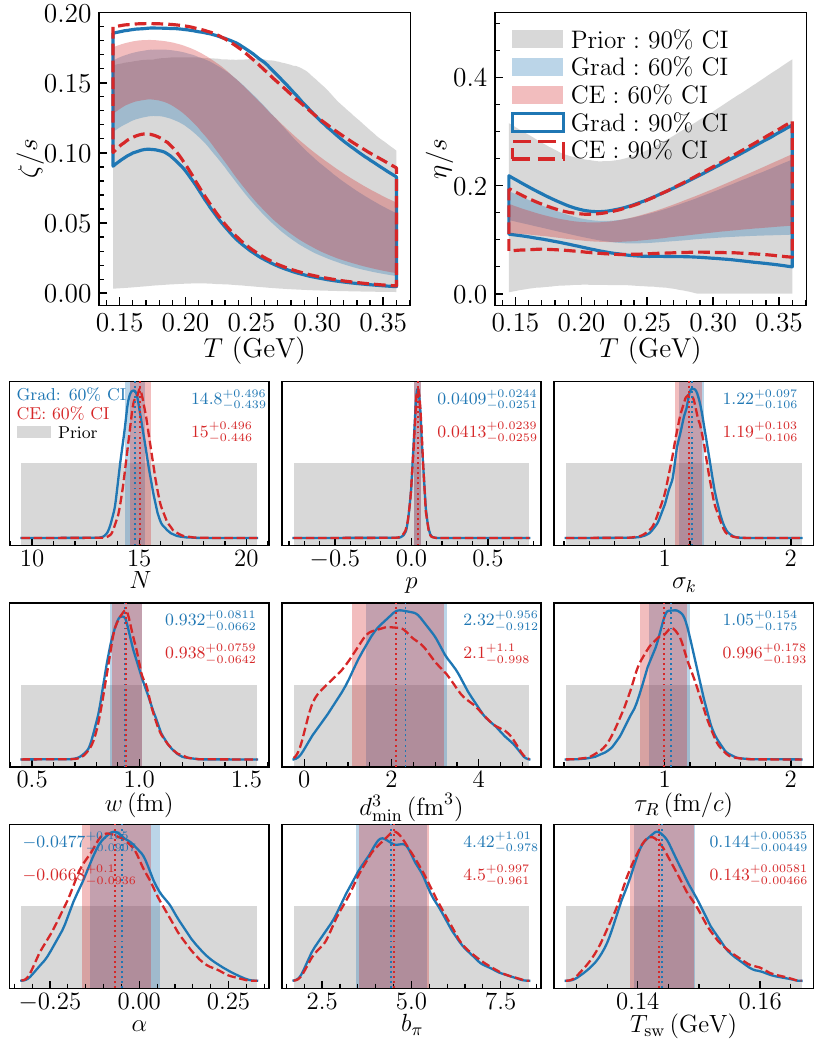}
  \vspace{-6mm}
  \captionof{figure}{Parameter posteriors from calibration to ALICE measurements using only $dN_{\text{ch}}/d\eta$, $dE_T/d\eta$, and $v_n\{2\}$ ($n=2,3,4$), without accounting for model discrepancy (w/o MD). The plot layout and legend are identical to those in Fig.~\ref{fig:param_woMD}. Good agreement between Grad and CE is observed for the specific viscosities and model parameters. However, the posteriors for $\zeta/s$, $d^3_{\min}$, $\alpha$, and (for Grad) $T_{\rm sw}$ shift significantly when all measurements are included in the inference (see Fig.~\ref{fig:param_woMD}), indicating a lack of robustness of the parameter estimates. For $\zeta/s$, posterior at low temperatures in Fig.~\ref{fig:param_woMD} shifts well outside the posterior band seen here.}
  \vspace{-2mm}
  \label{fig:param_woMD_subset}
\end{minipage}\hfill
\begin{minipage}[t]{0.47\textwidth}
  \centering
  \includegraphics[width=\linewidth]{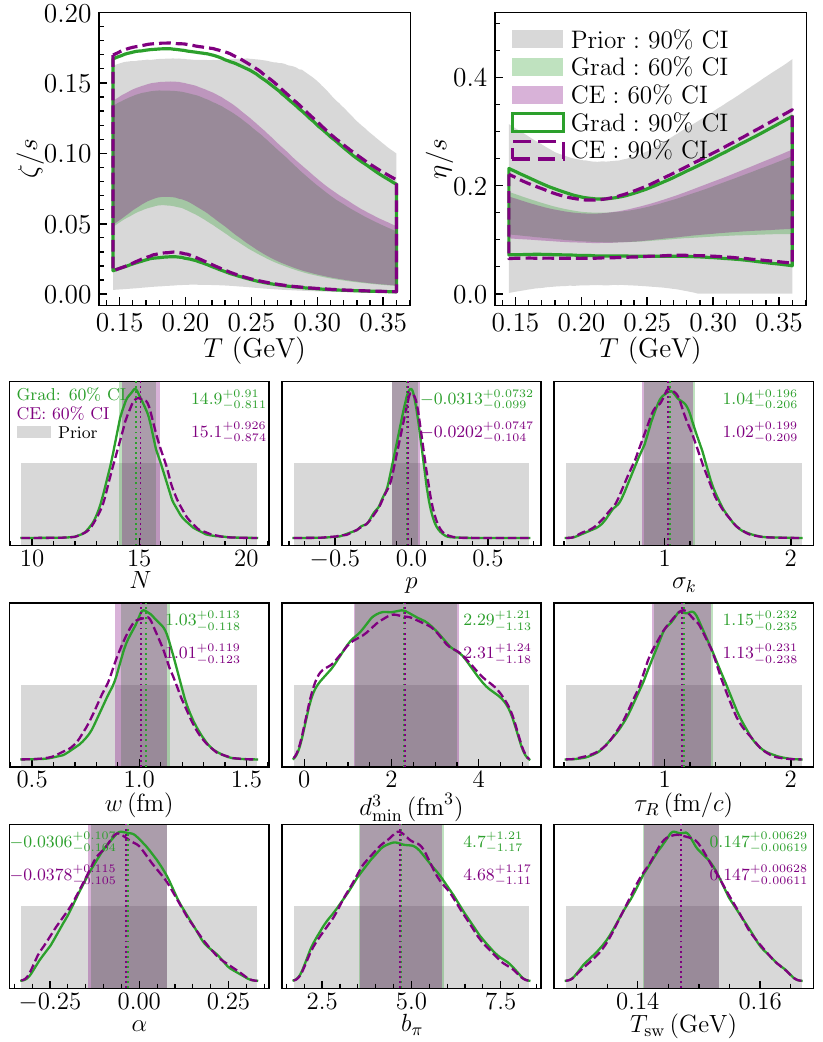}
  \vspace{-6mm}
  \captionof{figure}{Parameter posteriors from calibration to ALICE measurements using only $dN_{\text{ch}}/d\eta$, $dE_T/d\eta$, and $v_n\{2\}$ ($n=2,3,4$), with quantified model discrepancy (w/ MD). The plot layout and legend are identical to those in Fig.~\ref{fig:param_woMD}. Grad and CE posteriors agree across all parameters. Moreover, when all measurements are included (Fig.~\ref{fig:param_wMD}), the posteriors tighten without shifting (as occurred in the w/o MD case) indicating robust parameter inference.}
  \vspace{-2mm}
  \label{fig:param_wMD_subset}
\end{minipage}
\end{strip}

\begin{figure*}[t!]
    \centering
    \includegraphics[width=\linewidth]{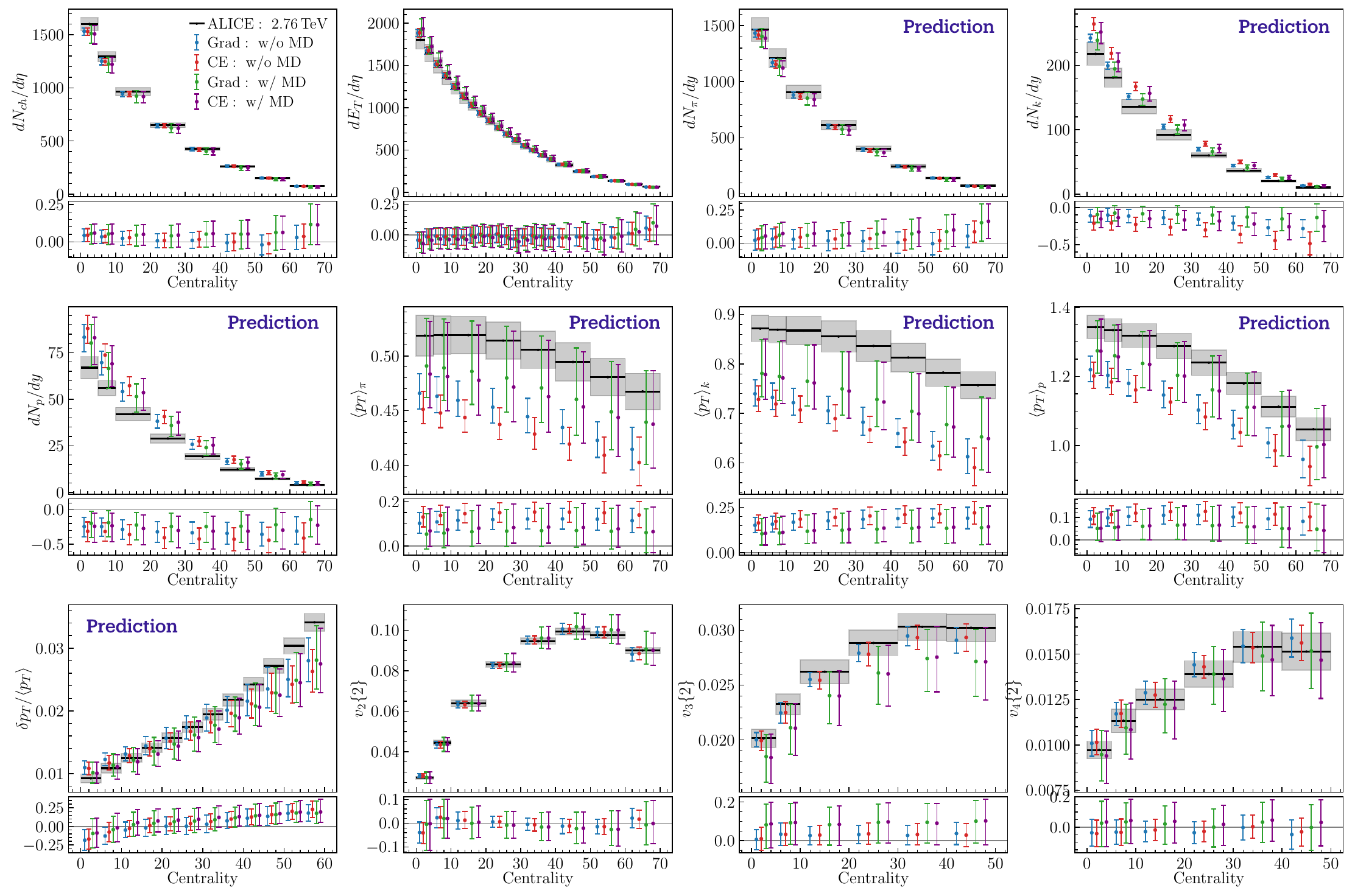}
    \vspace{-6mm}
    \caption{Model predictions obtained using the parameter posteriors in Figs.~\ref{fig:param_woMD_subset} and \ref{fig:param_wMD_subset} (calibrated only to $dN_{\text{ch}}/d\eta$, $dE_T/d\eta$, and $v_n\{2\}$ with $n=2,3,4$) compared with ALICE measurements (black). Observables not used in calibration are labeled ``Prediction'', and panels below each observable show the normalized discrepancies $(y_{\rm exp}-\eta_{\rm mod})/|\langle y_{\rm exp}\rangle|$. For observables included in the calibration, the w/o MD results agree more closely with the data than the w/ MD results for both Grad and CE. However, for the ``Prediction'' observables, the w/o MD predictions show large deviations from the data, consistently and well outside the predicted $68\%$\,CI for $\langle p_T \rangle_i,\ i\in \{\pi, k, p\}$, reflecting overfitting of the multistage model to the calibration data. The w/ MD results for these observables show comparatively better agreement with the data, albeit with large error bars.}
    \vspace{-2mm}
    \label{fig:obs_model_subset}
\end{figure*}

\clearpage
\bibliographystyle{elsarticle-num}
\bibliography{ref}

\end{document}